\documentclass[oneside,british]{amsart}
\usepackage[T1]{fontenc}
\usepackage[latin9]{inputenc}
\usepackage{amsthm}
\usepackage{amssymb}
\usepackage{esint}
\usepackage{xargs}[2008/03/08]

\makeatletter
\numberwithin{equation}{section}
\numberwithin{figure}{section}

\makeatother

\usepackage{babel}
\begin{document}

\title[Beyond Cumulated Gain and Average Precision]{Beyond Cumulated Gain and Average Precision: Including Willingness
and Expectation in the User Model}

\author{B. Piwowarski}

\address{Université Pierre et Marie Curie}

\email{benjamin@bpiwowar.net}

\author{G. Dupret}

\address{Yahoo! Labs USA}

\email{gdupret@yahoo-inc.com}

\author{M. Lalmas}

\address{Yahoo! Labs Barcelona}

\email{mounia@acm.org}
\begin{abstract}
In this paper, we define a new metric family based on two concepts:
The definition of the stopping criterion and the notion of satisfaction,
where the former depends on the willingness and expectation of a user
exploring search results. Both concepts have been discussed so far
in the IR literature, but we argue in this paper that defining a proper
single valued metric depends on merging them into a single conceptual
framework. 
\end{abstract}
\maketitle
\newcommandx\pr[2][usedefault, addprefix=\global, 1=]{p_{#1}\left(#2\right)}

\newcommandx\prc[3][usedefault, addprefix=\global, 1=]{\pr[#1]{#2\middle|#3}}
\global\long\def\waitExp{\mathcal{E}}
\global\long\def\waitWill{\mathcal{W}}

\global\long\def\Esp#1{\mathbb{E}\left[#1\right]}

\global\long\def\Espc#1#2{\mathbb{E}\left[#1\middle|#2\right]}

\global\long\def\satisfaction{\mathcal{S}}
\global\long\def\Nuggets{\mathcal{N}}
\global\long\def\stops#1{F#1}

\section{Introduction}

In the last few years, Information Retrieval~(IR) metrics have regained
attention within the IR community, motivated by concerns about the
faithfulness of the measures in IR, by trying to explicitly define
the user behaviour within an effectiveness measure~\cite{Robertson2008A-new-interpretation-of-average,Moffat2008Rank-biased-precision}.
Parallel to these efforts, the advantages of single valued metrics
have been argued since the very beginning of IR evaluation~\cite{Cooper1973On-selecting-a-measureII}.
Nonetheless, their interest lie in the fact that they provide an ordering
between systems, which is useful when one wants to compare many different
systems or to use machine learning when learning to rank. 

Developing a single valued measure that faithfully represents a system
is both a crucial and not fully resolved issue. In this paper, we
propose a single valued metric family relying on two factors, the
stopping criterion and the user satisfaction. Both factors are already
mentioned in the IR literature, but were surprisingly never explicitly
combined within a single metric. More importantly, with respect to
the stopping criterion, we propose a new rationale for its definition:
It should depend on both the \emph{willingness} of the user to pursue
browsing the search results and its \emph{expectation} to find new
relevant material. We analyse two recently proposed user models, one
for average precision~(AP) in~\cite{Robertson2008A-new-interpretation-of-average},
and Rank-Biased Precision~(RBP~\cite{Moffat2008Rank-biased-precision}),
that are representative instances of user models developed for the
two different different families of metrics proposed so far in IR~(one
based on precision and recall, the other on cumulated gain~\cite{Jarvelin2002Cumulated-Gain-based}),
in order to stimulate debate on the need of a new type of user models
for IR evaluation.

\section{The metric family}

\label{sec:sabor}

Following standard IR effectiveness measures, we assume that a user
browses the list of search results sequentially. We can imagine that
at each \emph{step~}(rank in the list) of the searching process,
we can measure whether the user will stop or continue, and in the
former case what is the \emph{degree of satisfaction} of the user.
The expected satisfaction over all possible stopping points defines
our metric. 

Modelling when the user stops is a key concept if we want to compute
a single valued measure~\cite{Cooper1973On-selecting-a-measureII}.
From a practical point of view, Cooper defined the stopping criteria
to be either a given rank or a given level of recall. In its interpretation
of AP, Robertson~\cite{Robertson2008A-new-interpretation-of-average}
describes the user model by specifying what is the probability that
a user stops to search \emph{exactly }at a given rank. Another approach
is that of~\cite{Moffat2008Rank-biased-precision} that estimates
the conditional probability that a user stops \emph{knowing }that
the user has continued until then. 

The conditional approach appears to be a more natural way to define
the fact that a user stops, since it is a local decision, i.e., we
do not have to integrate out of all possible user behaviours at previous
ranks. 

We suppose that stopping depends on two different but related factors:
How much effort~(e.g, in time or ranks%
\footnote{We favour the former, but in this paper consider only the latter to
analyse RBP and AP%
}) the user is \emph{willing} to spend before finding a new relevant
document, and how much effort the user \emph{expects} to spend before
finding a new relevant document. These are in turn influenced by numerous
factors, for example, how many relevant information the user has seen
so far, the effort the user has to spend to get to the next results,
how much more relevant information is present in the ranking, etc.
This view would account for the fact that having a result at rank
2 or 3 does not make a difference, to a certain extent, in the measured
user satisfaction, since users compensate for poor results~\cite{Smith2008User-adaptation}:
They are \emph{willing} to explore more when they do not find a relevant
document, until their \emph{expectation} of finding a relevant document
drops below a given threshold.

From a pragmatic point of view, the \emph{willingness} should depend
on how much more relevant documents the user expects to find, and
thus relates to the total number of relevant documents~(recall).
More precisely, the (effort) willingness should be a decreasing function
of a user's estimation of the number of relevant documents left in
the ranking. The \emph{expectation }of the number of ranks to browse
before finding relevant information should be evolving according the
current concentration of relevant~(and also novel) documents. Among
other features that could be explored, we believe the expectation
should also reflect the user belief in the performance of the search
engine~(which could in turn be used as a \emph{subjective} measure
of performance). 

Put together, and to relate it to the two metrics we analyse in this
paper, we should expect the user to stop more if the precision decreases~(expectation)
and when the perceived number of left relevant documents decreases~(willingness).
A proper definition of the user model is not discussed in this paper,
but this user model could be evaluated using e.g. query logs. 

Since we want to compute the expectation of satisfaction at the end
of a user's search, only when the user stops do we compute satisfaction~(or
more generally, a utility value). While satisfaction could be represented
in many ways, for normalisation purposes we assume that the satisfaction
$\satisfaction$ is a real value bounded by $0$ and $1$. Satisfaction
should take into account the effort of the user and the amount and
diversity of relevant documents found so far, and depends on the type
of query. For example, with navigational queries, one relevant document
is enough, hence the satisfaction should be (close to) maximum at
the first rank where a relevant document appears, whereas for informational
queries, satisfaction should be related to the precision at the rank
the user stops.

At  rank $k$, the user stops with a probability$\prc{\stops{=k}}{\stops{\ge k}}$
and the expected satisfaction is $\Espc{\satisfaction}{\stops{=k}}$.
If the user continues, its expected satisfaction is $\Espc{\satisfaction}{\stops{>k}}$.
This implies that we can define by induction the expected satisfaction
of a user, since the expectation $\Espc{\satisfaction}{\stops{\ge k}}$
is defined as $\prc{\stops{=k}}{\stops{\ge k}}\Espc{\satisfaction}{\stops{=k}}+\prc{\stops{\not=k}}{\stops{\ge k}}\Espc{\satisfaction}{\stops{>k}}$.
Denoting $p_{k}$ the probability $\prc{\stops{=k}}{\stops{\ge k}}$
and $s_{k}$ the expectation $\Espc{\satisfaction}{\stops{=k}}$,
the expected satisfaction can be written as a closed-form formula:

\begin{equation}
\Esp{\satisfaction}=\sum_{k=1}^{\infty}\left(\prod_{u=1}^{k-1}\left(1-p_{u}\right)\right)p_{k}s_{k}\label{eq:sabor}
\end{equation}
where both $p_{k}$ and $s_{k}$ have to be defined in order to obtain
a computable metric. In this paper we focus on analysing how they
are defined in the case of existing metrics, and analysing to which
extent they verify our requirements, namely that the stopping criterion
should reflect both willingness and expectation, and that the satisfaction
at a given rank measures how satisfied~(in average) a user is when
stopping at a given rank.

\section{Discussion and Comparison with existing metrics}

It is instructive to see how AP and RBP are interpreted within our
proposed approach. In order to do so, we used the most straightforward
way to define the probabilities $p_{k}$ and the expected satisfaction
$s_{k}$ in both: For AP we followed~\cite{Robertson2008A-new-interpretation-of-average}
and re-interpreted the stopping criterion. RBP~\cite{Moffat2008Rank-biased-precision}
is an instance of Eq.~(\ref{eq:sabor}). 

An underlying user model can be defined for AP~\cite{Robertson2008A-new-interpretation-of-average}:
A user stops exactly at a given rank with a uniform probability. Re-interpreting
AP in terms of Eq.~(\ref{eq:sabor}), we offer a slightly different~(but
equivalent) interpretation: If reaching rank $k$, the user always
continue if the document is not relevant, and stops with a probability
which is inversely proportional to the number of relevant documents
on or after rank $k$ otherwise. This means that the AP stopping criteria
looks even more reasonable than what have been stated by Robertson~\cite{Robertson2008A-new-interpretation-of-average}
since a user is more likely to stop when finding more relevant documents.
The AP user's model thus relies on one of the two factors we defined,
the willingness, but ignore the expectation factor. With respect to
the expected satisfaction, it is simply defined as the precision at
the given rank which implies AP is more targeted towards informational
needs.

The rank-biased precision~(RBP) metric proposed by Moffat and Zobel~\cite{Moffat2008Rank-biased-precision}
defines $p_{k}$ as a constant and $s_{k}$ as the gain of a document.
This has nonetheless two shortcomings: First, the constant of probability
of stopping implies in our opinion a worse user model than that of
AP, since none of the two factors are taken into account. Second,
the assumption that the gain $g_{k}$ is associated only with the
document at rank $k$ implies that the user satisfaction only depends
on the document where the user stops, which does not comply with our
interpretation of $s_{k}$ as an overall measure of satisfaction.
Whereas this might be reasonable for navigational needs, for informational
ones either the discount should not be interpreted as the probability
of stopping at this rank, or the gain should not be interpreted as
a referring only to the document where the user stops.

\section{Conclusion}

We presented a new family of metrics based on the user's stopping
probability and expected satisfaction~(if stopping) at each rank.
For the former, we propose to take into account both the user \emph{willingness}
of continuing searching and \emph{expectation} of finding new relevant
documents. We believe that these two factors are a key to provide
meaningful evaluation measures.

We have analysed two representative metrics in IR, AP and RBP~\cite{Moffat2008Rank-biased-precision}.
With respect to the latter, we have shown that interpreting user satisfaction
as precision at a given rank, the user stopping behaviour respect
our willingness requirement but ignores the expectation of the user
to find new relevant documents. With respect to the former, we have
shown that both the user model and the definition of satisfaction
did not match our expectations. This has implications for works trying
to define a user model for CG based metrics. In particular, if the
gain is to be interpreted as in~\cite{Jarvelin2002Cumulated-Gain-based},
then the discount factor should not be interpreted as the probability
of a user stopping at this rank.

Our future work is to work on both the stopping criterion and the
definition of the satisfaction.

\bibliographystyle{abbrv}
\bibliography{sabor}

\end{document}